# Towards the classification of Self-Sovereign Identity properties

**ŠPELA ČUČKO[1], ŠEILA BEĆIROVIĆ[2], AIDA KAMIŠALIĆ[1], SAŠA MRDOVIĆ[2], AND MUHAMED TURKANOVIĆ.[1]**
[1]Faculty of Electrical Engineering and Computer Science, University of Maribor, Koroška cesta 46, 2000 Maribor, Slovenia
[2]Faculty of Electrical Engineering Sarajevo, University of Sarajevo, Zmaja od Bosne bb, Sarajevo 71000, Bosnia & Herzegovina

Corresponding author: Špela Čučko (e-mail: spela.cucko@um.si).

This work was supported by the Slovenian Research Agency (Research Core Funding) under Grant P2-00577, and by the European Union's Horizon 2020 research and innovation program under Grant Agreement No 870635 (DE4A).

**ABSTRACT** Self-Sovereign Identity (SSI) is a novel and emerging, decentralized digital identity approach that enables entities to control and manage their digital identifiers and associated identity data fully while enhancing trust, privacy, security, and the many other properties identified and analyzed in this paper. The paper provides an overview and classification of the SSI properties, focusing on an in-depth analysis, furthermore, presenting a comprehensive collection of SSI properties that are important for the implementation of the SSI system. In addition, it explores the general SSI process flow, and highlights the steps in which individual properties are important. After the initial purification and classification phase, we then validated properties among experts in the field of Decentralized and Self-Sovereign Identity Management using an online questionnaire, which resulted in a final set of classified and verified SSI properties. The results can be used for further work on definition and standardization of the SSI field.

## I. INTRODUCTION

Self-Sovereign Identity (SSI) is an emerging, decentralized identity concept that enables entities (e.g., individuals, organizations, and things) to control and manage their digital identity fully [1] without dependency on any external authority, eliminating a single point of failure, while enhancing trust, privacy, security [2], and many other properties, such as transparency, persistence, interoperability, minimalization, etc. The concept is gaining momentum with the rise of the Blockchain technology. Its potential in the field of Identity Management was first recognized in 2015, when the Internet Identity Workshop (IIW) started a discussion about blockchain identity [3]. This was followed by other initiatives, including Allen [4], who proposed ten guiding principles of SSI. However, SSI is still in its infancy without a consensus on the exact definition, without precisely defined architecture or implementation. Thus, various aspects of SSI have been examined in the literature, including initiatives to describe and formally define the concept [5], essential architectural components [6], [7], underlying technology [8] process flows [5], and principles/properties [4], [5], [9], [10] important for its implementation. Unfortunately, there are some inconsistencies among the identified properties, their naming, and the definitions of various authors. Several sets of SSI properties have been defined, but some overlap. Therefore, the aim of this research was to collect SSI properties defined in the literature, analyze them in detail, classify them into meaningful groups, and, finally, validate the optimized, i.e., final set of properties and the classification, by surveying experts in the field of Decentralized and Self-Sovereign Identity Management. To the best of our knowledge, this is the first such attempt to address the properties of SSI comprehensively, which, in our belief, is a crucial part of any young research field which has still to evolve and requires solid foundations to build (i.e., research) upon.

All in all, this paper combines deductive and inductive approaches while exploring fundamental properties that an SSI should have, and strives toward their classification. Therefore, the contribution of this paper is as follows: (i) An overview and analysis of SSI properties from the literature, (ii) A comprehensive collection of SSI properties, (iii) The classification of the properties, (iv) The validation of the final set of SSI properties.

The remainder of this paper is structured as follows. The general concepts of SSI, its properties, and process flow, are outlined in Section II. Furthermore, related work regarding SSI properties is presented, while differences to our work are also highlighted. Section III consists of multiple subsections, and discusses the conducted analysis. Subsection III-A offers a list of collected properties that are classified in Subsection III-B and connected to steps in the general SSI process flow in Subsection III-C. Additionally, Subsection III-D analyzes the importance of individual properties for various roles, and Subsection III-E highlights the properties that can be satisfied by utilizing Blockchain technology. Section IV validates the final set of properties and proposed classification. The research methodology is described in Subsection IV-A, and the respective results are presented and discussed in Subsection IV-B. The penultimate Section V discusses properties and represents the final state. It reflects the findings of the validation phase and summarizes the final properties, their definitions, and classification. Last but not least, Section VI presents conclusions and future work.

## II. BACKGROUND AND RELATED WORK

Self-Sovereign Identity is a decentralized identity approach that provides the means for digital identification and authentication, allowing entities (e.g., individuals/users, organizations, and things) to control their digital identity fully. It presents the next step in the field of Identity Management (IdM) that is moving away from traditional, centralized, and federated identity systems that rely on external entities for identity provisioning and management. Current systems require the user either to create an account with each service provider (SP), or to create an identity, i.e., identifier and credential, with one of the Identity Providers (IdP), and then use the latter to authenticate and access various services within a federation of SPs. Identifiers and identity attributes in both models are stored centrally under the control and management of an IdP or SP who are central in digital interactions, while their internal policies influence the privacy and security of collected data greatly. On the other hand, SSI allows entities to generate, manage, store, and control their identity without the intervention of external parties, eliminating intermediaries in digital interactions, and preventing scenarios where third-party data requestors obtain data directly from third-party data providers.

SSI has emerged with the rise of blockchain technology that can facilitate some desired features of digital identity, minimize, or even eliminate fully, reliance on any external authority, and solve some problems that traditional identity management systems entail. As identifiers and associated identity data are no longer stored in centralized third-party repositories, eliminating a single point of failure, reducing the threat to privacy, enhancing security, and minimizing vulnerabilities connected to personal data misuse, data breaches, and identity-related cybercrimes [11]. Furthermore, SSI is user-centric [6], presenting a shift of power and control from central authorities to decentralized entities such as users, i.e., identity holders, who must be central to the administration of their own identity and information flow during digital interactions [12], [13] and are responsible for storing their credentials in user agents, i.e., wallets. SSI enables an exchange of claims and credentials without an intermediary, allowing users to attain verifiable credentials from third-party issuers, and/or make assertions about themselves and present them to the relying party, i.e., verifiers, requesting proof of identity. Thus, claims and credentials must be verifiable to be trusted.

Some of the main objectives of the SSI system are to (i) Enhance a user's full control in digital interactions involving the exchange of personally identifiable information (PII), (ii) Ensure security and privacy, as well as enable partial, but yet verifiable, disclosure of information, (iii) Ensure that PII is shared only with the consent of its bearer, (iv) Prevent altering the data, and (v) Ensure the adequacy of data that can be trusted and verified by the relying party [14].

Various aspects of SSI have been examined in the literature, including the properties of SSI, reflecting the main objectives mentioned above. The reason for this is certainly the fact that presenting the properties of a concept is one of the best possible ways to describe and encompass it objectively. If generalized, research addressing the SSI properties can be differentiated into three groups, while each is related to our study to a certain extent. Some are related directly (b), while others are related indirectly (a, c). (a) First of all, there is some groundwork where researchers propose SSI principles/properties. (b) Secondly, there are studies classifying properties or reviewing, analyzing, and commenting on proposed properties, either rejecting or confirming them, or suggesting new ones. (c) Finally, some studies use the proposed properties as evaluation criteria to verify that the identity system is indeed SSI. The second group (b) is related to our study the most. However, neither has validated properties nor their classification among experts in the field of Decentralized and Self-Sovereign Identity Management.

Allen [4] has proposed ten guiding principles of the SSI, laying out the foundation for implementation of the concept, stating that the key properties of SSI system are Existence, Control, Access, Transparency, Persistence, Portability, Interoperability, Consent, Minimization, and Protection. The aforementioned SSI properties can be connected to steps in the general SSI process flow that is presented in Figure 1. In addition, the main actors involved in the process can also be observed, and include user (identity holder), issuer (identity attributes provider), and verifier (service provider) that interact with each other. All actors should be able to generate and manage multiple unique decentralized identifiers (DID) independently of any third party [15] (Existence) and acquire identity attributes (verifiable claims and credentials, VCs) from third-party issuers (Access).

Moreover, identifiers (Existence) and associated personal data can be stored and managed securely and autonomously (Control) by identity subjects, while attained attestations can be presented freely when proof of identity is required (Control, Consent, Minimalization) [11]. Furthermore, with the use of decentralized technologies and cryptographic primitives, security, and privacy (Protection) are enhanced, and Transparency and data Minimalization can be achieved.

The Sovrin Foundation summarized and grouped Allen's principles into three sections (i) Security (Protection, Persistence, Minimalization), (ii) Controllability (Existence, Persistence, Control, Consent), and (iii) Portability (Interoperability, Transparency, Access). Highlighting that, in essence, (i) Identity information must be kept secure, while (ii) The users remain in control of their data and must be able to determine who can access it, moreover, (iii) Identity must be available, widely usable, and portable, not tied to a single identity provider [16]. The Sovrin Foundation [10] has also listed twelve foundational SSI principles. As presented in Table 2, some proposed properties are similar to Allens. In addition, they have extended the principle Control by adding the possibility of employing and/or delegating control to agents and guardians of an entities' choice, while also highlighting the importance of (i) Decentralization, (ii) Equity and Inclusion, (iii) Usability, Accessibility and Consistency, (iv) Verifiability and Authenticity.

Stokkink and Pouwelse [17] concluded that most of the properties proposed by Allen can be achieved intrinsically by leveraging a personalized blockchain structure. However, some open challenges concerning Portability, Interoperability, Minimalization, and Protection remain, and refer to the claim structures. Meanwhile, the authors have also highlighted the need for the additional requirement for SSI, as claims need to be provable in order to be valid, which coincides with the property Verifiability and Authenticity already mentioned by Sovrin.

Toth and Anderson-Priddy [9] reviewed and evaluated the work of Cameron, Allen, W3C Verifiable Claims Working Group, and Sovrin. They validated nine proposed properties and suggested five additional properties, namely, (i) Usability, (ii) Counterfeit Prevention, (iii) Identity Verification, (iv) Identity Assurance, and (v) Secure Transactions, addressing situations dealing with the loss of digital identities. On the other hand, they argue that Existence, Transparency, and Protection proposed by Allen require further discussion, and should be set aside, as Existence is self-evident, Transparency might not always be possible, and Protection involves critical and demanding policy issues that should be addressed in more detail. To reason and validate the final set of properties, they were applied to SSI architecture.

Furthermore, Ferdous et al. [5] examined SSI properties in detail by analyzing existing definitions critically and extracting properties aiming to propose a formal definition of the concept. They classified properties into five categories, (i) Foundational (Existence, Autonomy, Ownership, Access, Single source), (ii) Security (Protection, Availability, Persistence), (iii) Controllability (Choosability, Disclosure, Consent), (iv) Flexibility (Portability, Interoperability, Minimization), and (v) Sustainability (Transparency, Standard, Cost), providing their taxonomy. In addition, the authors have presented several use cases involving SSI, and highlighted the essential life cycles of an identity management system.

The aforementioned properties can be viewed as a set of requirements that SSI systems should achieve. Therefore, they can be used as evaluation criteria for determining if an identity system is self-sovereign or not. Thus, as already mentioned, some work regarding the assessment of digital identity solutions exists in the literature [5], [8], [17]–[19]. Soltani et al. [18] assessed their proposed client onboarding framework (KYC2) against various criteria, including SSI principles, by Allen [4]. Bokkem et al. [19] conducted a comparative study, reviewing and evaluating several blockchains and non-blockchain SSI solutions based on properties described by Allen [4], accompanied by the Provability property proposed by Stokkink and Pouwelse [17]. They concluded that Blockchain technology is a good foundation for SSI implementation, but is not required explicitly. Systems utilizing Blockchain technology, however, meet more SSI properties. Similarly, Ferdous et al. [5] investigated white papers and technical documents of four SSI systems, namely uPort, Jolocom, Sovrin, and Blockcerts, analyzing if they satisfy different SSI properties. Ferdous et al. and Bokkem et al. [5], [19] noted that, according to the evaluation criteria, i.e., SSI properties, identity systems characterized as SSI, in most cases, do not satisfy all the identified properties fully, or the latter is not clear from the documentation. However, according to Bokkem et al. [19], some meet all the criteria [4], [17], including Sora, ShoCard, SelfKey, LifeID.

### III. ANALYSIS

In contrast to the aforementioned studies based on research by a handful of individuals (usually researchers), we focused on the opinion of experts in the field of Decentralized and Self-Sovereign Identity Management, occupying different positions within different domain areas. We wanted to gain a broader view of the perception of the concept, its importance, and determine which properties are the most important, even mandatory, for the implementation of SSI. Therefore, we have analyzed and classified identified properties, explored properties (in relation to the (a) General SSI process flow, (b) Various roles, and (c) Blockchain technology), and conducted a questionnaire with the objective to (i) Investigate the perceived level of importance of each identified SSI property, (ii) Determine a set of the least and the most important properties, i.e., non-negotiables, and (iii) Verify the appropriate classification/grouping of properties. In addition, most of the studies presented in the previous section, use, critique, or extend Allen's principles [4] that

**IEEE** Access*

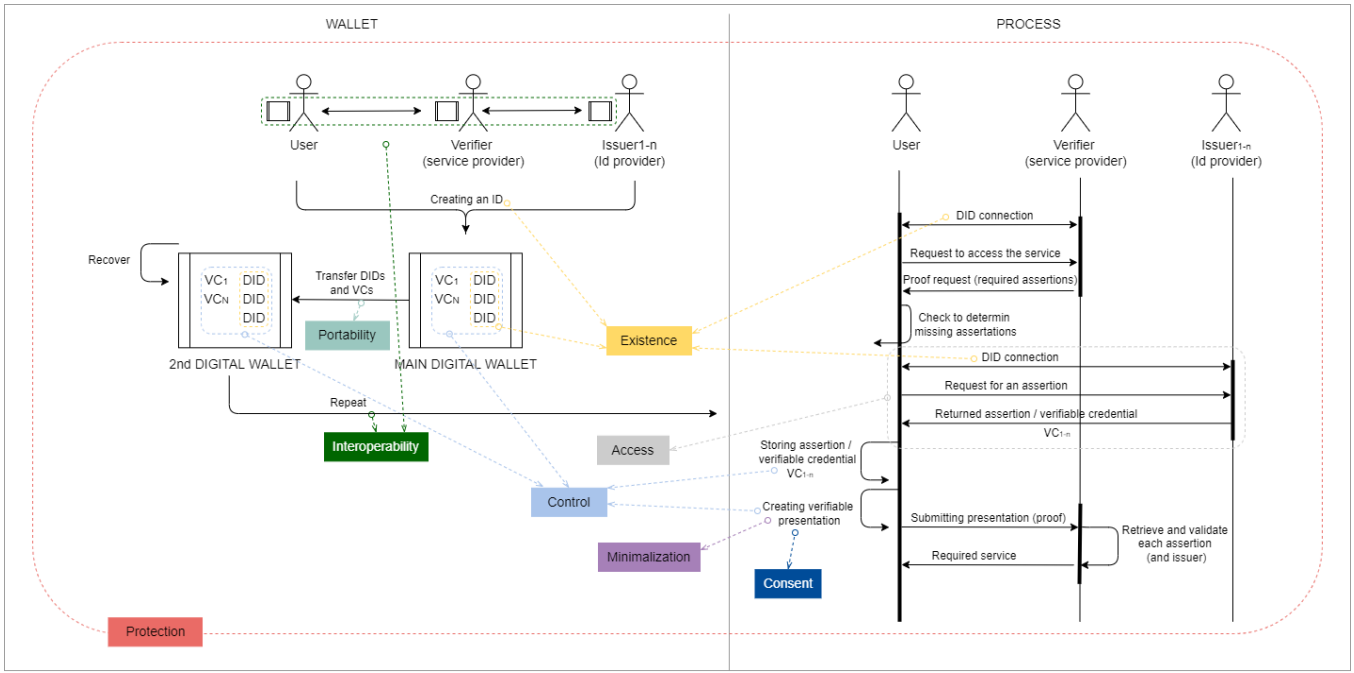

**FIGURE 1.** The general SSI process flow and connected properties proposed by Allen [4].

present their research starting point. In contrast, Ferdous [5] has derived properties from records and unofficial definitions of SSI published mainly on forums.

On the other hand, our study takes into account already defined properties. We optimize them and use them in the questionnaire, consequently dealing with a larger set of properties.

### A. PROPERTIES

The literature review and our preliminary research showed inconsistencies among the identified properties, their naming, and the definitions of various authors. Therefore, we gathered a list of the predefined properties presented in the previous Section. For each, we collected all the definitions, analyzed them in detail, and extracted key features. Afterwards, some properties were grouped, some were eliminated, and some were added, as we observed that properties of different authors overlapped. The latter is visualized in Figure III-A, where similarities and differences can be observed in naming between identified sets of properties. The overlapping properties that describe the same thing and are named differently were combined and treated as one, as shown in the Figure with the arrows, while we have adopted more appropriate naming. Thus, the final set of 18 properties is listed in Figure III-A, and their definitions can be found in Table 1.

**Methodology:** The analysis and grouping of properties took place in several iterations. The steps are listed below.

1) Identification and collection of properties from various sources.

2) Recording of properties and their definitions according to the author, which were, furthermore, labeled accordingly (1. Labeling).

3) Comparison and grouping of properties according to their name. Properties with the same naming and similar definition were combined (1. Iteration of grouping).

4) Denoting key constructs (features) that characterize the essence of a property and recording connections between them (2. Labeling).

5) Comparison and grouping of properties according to their definitions and key constructs. Properties that had the same or similar definition but were named differently were grouped (2. Iteration of grouping).

6) Grouping, eliminating, and adding properties after discussing our reasoning with experts. To verify our decisions and finalize the list of properties (used in the questionnaire), interviews with experts were conducted, where the reasoning behind our decisions was discussed (3. Iteration of grouping).

7) Designing the final list and definitions for each property separately.

**Results:** Table 2 shows the results of the analysis, including the similarities and differences in naming between defined sets of properties. We have combined the overlapping ones. Therefore, properties that describe the same thing and are named differently were treated as one, while we adopted more appropriate naming. Each Table row represents one property according to the similarity of the collected definitions, while differences in naming can be observed between

IEEE Access

**TABLE 1.** Self-Sovereign Identity properties and their definitions.

| Property | Definition* |
|---|---|
| Existence and Representation | Entities must have an independent existence. They should be able to create as many identities as required without the intervention of a third party. |
| Decentralization and Autonomy | Entities must have full autonomy over their identity data without relying on any third party (centralized system). They should be responsible for managing all operations related to their identity and data (creating, storing, updating, sharing, removing). |
| Ownership and Control | Entities must own and control their digital identities and the involved data fully (e.g., self-asserted claims or claims provided by third parties, identifiers, encryption keys). They should be able to control the usage/sharing of their identity data and delegating control to autonomous agents and/or guardians of their choice. |
| Privacy and Minimal Disclosure | Entities should be able to protect their privacy by utilizing selective disclosure and data minimization. They should be able to disclose the minimum amount of identity data required for any particular interaction. |
| Single source | Entities should be the single source of truth regarding their identities. They should be able to create self-asserted claims, accumulate claims from third parties, and distribute them when required. Third parties should not be able to exchange entities' data without their knowledge and consent. |
| Consent | Entities should be able to give deliberate and well-understood consent for the usage/sharing of their identity data (e.g., consenting to accept data, related to their identity). |
| Security and Protection | Digital identity should be secure and well protected with reliable cryptographic mechanisms. Entities must be authenticated and authorized properly prior, to be able to use their digital identity. Any identity information must be transmitted/transferred via a secure channel to prevent cyber attacks. |
| Verifiability and Authenticity | Entities must be able to prove their identity reliably. They must provide verifiable proof of authenticity of digital identity data. Relying parties should be able to verify that digital identities are controlled by their owners and have not been tampered with. |
| Accessibility and Availability | Entities must have unrestricted access to their identity information. They must be able to retrieve claims and assertions (self-asserted or provided by a third party) that constitute their identity, and must be accessible and available from different platforms when required. |
| Recoverability | The identity must be robust enough to be recoverable. |
| Usability and User Experience | The usability of agents and other identity system components should be maximized. User interfaces should allow entities to control, manage, and use their identities intuitively, reliably, and effectively. It should offer a consistent user experience, hide underlying complexity, and should be easy to use. |
| Transparency | The identity system and algorithms must be transparent enough for every involved entity. They should be free, open-source, well-known, and independent of any particular architecture. Entities should be well aware of all their partial identities and their corresponding interactions. |
| Standard | Identities must be based on open Standards to ensure maximal portability, interoperability, and persistence. Entities should be represented, exchanged, secured, protected, and verified using open, public, and royalty-free Standards. |
| Persistence | Identities must be persistent, and should exist for at least as long as it is required by their owner. Longevity and the dynamic nature require firm separation between identity and its claims that can be modified or removed as appropriate. |
| Portability | Identities must be portable. Entities should be able to move or transfer their identity data to agents or systems of their choice securely. Portability ensures entities' control over their data and improves persistence over time. |
| Interoperability | Identities must be as widely usable/available as possible, and not limited to a specific domain. Global identities might increase persistence and identity autonomy. |
| Compatibility with legacy systems | Identity should be backward compatible with legacy identity systems to ensure quicker acceptance. |
| Cost | The cost of identity creation, management, and adoption should be minimized. |

* Due to their use in the questionnaire, we tried to shorten the definitions as much as possible, but we tried to keep their essence.

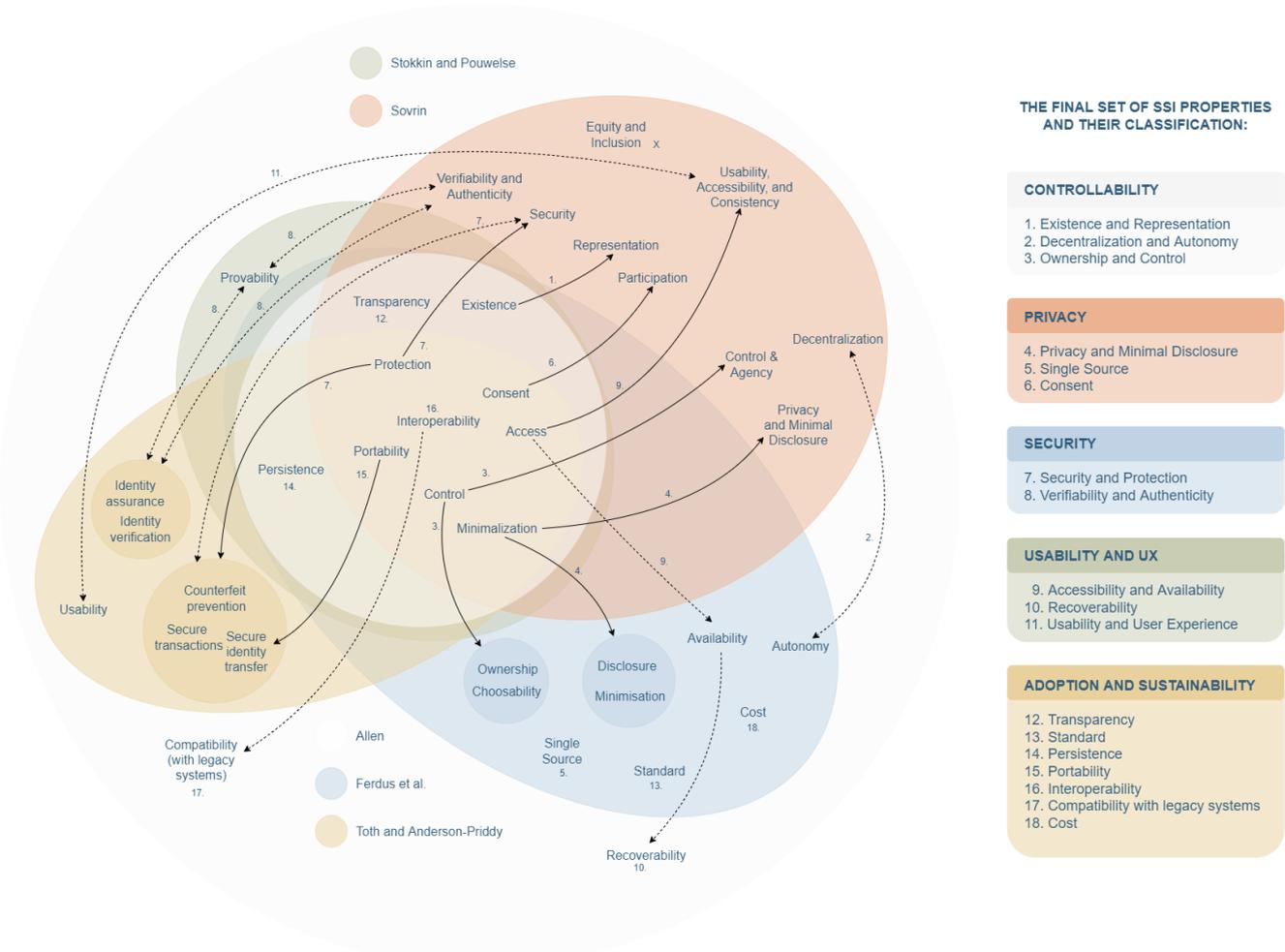

**FIGURE 2.** Various sets of SSI properties, defined by different authors on the left. The final set of properties and their classification on the right.

different authors. Consequently, properties (i) Existence and Representation, (ii) Ownership and Control, (iii) Accessibility and Availability, (iv) Security and Protection, and (v) Decentralization and Autonomy were combined. On the other hand, the property Equity and Inclusion was eliminated, while the properties Recoverability and Compatibility with legacy systems were added. Thus, our rationale is explained below.

Recoverability was derived from Availability, defined by Ferdous et al. [5], since we believe it is the important property itself, and has already been recognized as a crucial challenge that needs to be tackled [6], [11], [20]–[22]. While traditional, central, and federated identity models provide key-management approaches based on a trusted third party, under the SSI model the responsibility for key and wallets' management is under the control of identity holders [20], allowing self-sovereignty, but poses risks and challenges that can influence user experience and the adoption of SSI solutions greatly. With control, a lot of responsibility is transferred to users, who cannot relying on the support of central authorities in case of problems (e.g., forgotten private keys, phone loss, digital wallet vulnerability, etc.). Therefore, one of the key challenges is related to the development of appropriate and effective mechanisms for managing and recovering decentralized identities, as the lack of appropriate protocols for their management and recovery can lead to vulnerability, data loss, and fraud [11], [20].

Property Compatibility with legacy systems was derived from property Interoperability, defined by Ferdous et al. [5], addressing compatibility with existing government systems, e.g., Federal Public Key Infrastructures (FPKI). We believe legacy systems will become obsolete and be replaced gradually by SSI systems, but their compatibility with SSI is more convenient in the early stages of adoption. Therefore, the mentioned properties should be separated, as properties are not contingent on each other, while carrying different levels of importance. However, both can facilitate wider acceptance.

We have explained only the reasons for adding new properties. The remaining properties are presented in Table

1, and will not be addressed further at this point since they have been analyzed in detail by other authors and are quite self-explanatory.

### B. CLASSIFICATION

Existing classifications [5], [16] offer different views on SSI properties. However, they do not cover the entire set of properties identified by our analysis, so they should be adjusted accordingly. Additionally, the Foundational category proposed by Ferdus et al. includes properties that we believe should, initially, be redistributed among other defined categories and highlighted as key properties.

Hence, after the analysis, eighteen properties were obtained and classified into five categories, namely, (i) **Controllability**, (ii) **Privacy**, (iii) **Security**, (iv) **Usability and User Experience** (UX), and (v) **Adoption and Sustainability**. The properties belonging to each category are presented in Table 3. (i) The category **Controllability** combines properties that allow entities to gain control over their identity, and includes the following: Existence and Representation, Decentralization and Autonomy, Ownership and Control. The latter is essential, as SSI is a decentralized identity approach that enables entities to control their digital identities fully without reliance on any external authority. (ii) Properties grouped under the **Privacy** category allow individuals to maintain privacy while interacting with third parties over the internet by providing or disclosing the minimum identity information required for specific interaction. Meanwhile, the latter can only be shared with consent from its identity holder/subject. Therefore, it includes the properties Privacy and Minimal Disclosure, Single source, and Consent. (iii) The properties grouped in the **Security** category are dealing with the security of identity data. The category focuses mainly on authentication and authorization, while providing security in every step of digital interaction, dealing with identity data. It includes Security and Protection, and Verifiability and Authenticity. (iv) **Usability and User Experience** address mainly user interfaces, e.g., agents, their design, usability, availability, accessibility, and ease of use. The category includes properties that can affect the differentiation between successful and unsuccessful systems significantly according to users' experience. (v) The properties group under the category **Adoption and Sustainability** can ensure quicker and wider acceptance of SSI, since Transparency and Standard instill trust in people while ensuring Portability, Interoperability, and Persistence. Minimal cost and Compatibility with legacy systems are convenient and desirable, especially in the early stages of adoption.

### C. PROCESS FLOW

In addition to the identified properties and their classification, we have found some use cases of SSI applied in various domains [5], [23]. By analyzing and observing them, we have noticed that the process can be generalized. Moreover, properties can be connected to specific steps in the process flow.

Therefore, in this Section, we explore the general SSI process flow in terms of (i) Identifier (DID) generation [24], (ii) Acquisition of verifiable credentials (VCs) from identity issuers [25], (iii) Storage of VCs, and (iv) Interaction with verifiers through verifiable presentations (VPs) to determine in which steps the identified properties are paramount. The process and connected properties are presented in Figure 3.

Each interaction requires establishing a pairwise, secure DID connection between interacting parties. Therefore, DID creation is a crucial step, as DIDs are identifiers, enabling verifiable decentralized identity, empowering entities (e.g., individuals, organizations, and things) with **Existence and Representation** while providing **Verifiability and Authenticity**. Afterwards, the user, i.e., the identity holder, can either (i) Request to access the service, or (ii) Request an assertion (a VC consisting of one or several claims) from identity issuers. (i) When acquiring a service the proof request is indeed needed, allowing the service provider to proceed with identification and verification. If the identity holder has all the required credentials, he/she can proceed with the process by submitting a VP that enhances privacy and allow users to disclose only the minimum amount of identity data required for the interaction. Hence, VPs facilitate **Ownership and Control**, **Privacy and Minimal Disclosure**, **Consent**, and are retrieved and validated by the service provider, where thorough **Verifiability and Authenticity** are required. (ii) Otherwise, he/she must obtain the appropriate assertions (VCs) from trustworthy issuers. VCs are then stored in the user's digital wallet, i.e., agent, along with identifiers, and can be presented as needed, meanwhile, remaining under the user's control, reinforcing autonomy and ownership of the identity and associated identity data. Storing identity data by users themselves largely preserves the following properties: **Ownership and Control**, **Decentralization and Autonomy**, and enables **Accessibility and Availability**, allowing unrestricted access and control of its own data. The **Availability** must also be ensured when interacting with issuers in the process of obtaining assertions.

Transferring identity data from one wallet or device to another enables **Portability**, while repeating the entire process of obtaining credentials and accessing services from another wallet or a device endows **Interoperability**. That is also crucial for the interaction between different digital wallets and agents in the possession of each entity.

In the event of problems, such as forgotten private keys, phone loss, digital wallet vulnerability, etc., **Recoverability** enables users to recover identity data successfully without having to reacquire previously obtained credentials.

**Security and Protection**, **Usability**, and a good **User Experience** must be ensured throughout the entire SSI process.

On the other hand, some properties cannot be applied to a specific process step, so there is a noticeable demarcation between process properties and general properties that include **Transparency**, **Standard**, **Persistence**, **Compatibility with legacy systems**, and **Cost**. Thus, general process properties are omitted from Figure 3.

**TABLE 2.** Comparison of identified properties in various sources.

| Allen [4] | Sovrin [10] | Toth and Anderson-Priddy [9] | Stokkink and Pouwelse [17] | Ferdous et al. [5] | Final set of properties |
|---|---|---|---|---|---|
| Existence | Representation |  | Existence* | Existence | Existence (and Representation) |
| Control | Control and Agency | Control* | Control* | Ownership Choosability | Ownership and Control |
| Access | Usability, Accessibility, and Consistency | Access* | Access* | Access Availability | Accessibility and Availability |
| Transparency | Transparency |  | Transparency* | Transparency | Transparency |
| Persistence |  | Persistence* | Persistence* | Persistence | Persistence |
| Portability | Portability | Portability* | Portability* | Portability | Portability |
| Interoperability | Interoperability | Interoperability* | Interoperability* | Interoperability | Interoperability |
| Consent | Participation | Consent* | Consent* | Consent | Consent |
| Minimalization | Privacy and Minimal Disclosure | Minimalization* | Minimalization* | Disclosure, Minimization | Privacy and Minimal Disclosure |
| Protection | Security | Secure identity transfer, Secure transactions, Counterfeit prevention | Protection* | Protection | Security and Protection |
|  | Decentralization |  |  | Autonomy | Decentralization and Autonomy |
|  | Equity and Inclusion |  |  |  |  |
|  | Verifiability and Authenticity | Identity verification, Identity assurance | Provability |  | Verifiability and Authenticity |
|  | Usability, Accessibility, and Consistency | Usability |  |  | Usability and User Experience |
|  |  |  |  | Single Source | Single source |
|  |  |  |  | Standard | Standard |
|  |  |  |  | Cost | Cost |
|  |  |  |  |  | Recoverability |
|  |  |  |  |  | Compatibility (with legacy systems) |

\* = Coincides with Allen's principle [4]. Meaning that the same naming and definition were used.

**TABLE 3.** SSI properties' classification.

| Controllability | Privacy | Security | Usability and User Experience | Adoption and Sustainability |
|---|---|---|---|---|
| Existence and Representation | Privacy and Minimal Disclosure | Security and Protection | Accessibility and Availability | Transparency |
| Decentralization and Autonomy | Single source | Verifiability and Authenticity | Recoverability | Standard |
| Ownership and Control | Consent | | Usability and User Experience | Persistence |
| | | | | Portability |
| | | | | Interoperability |
| | | | | Compatibility with legacy systems |
| | | | | Cost |

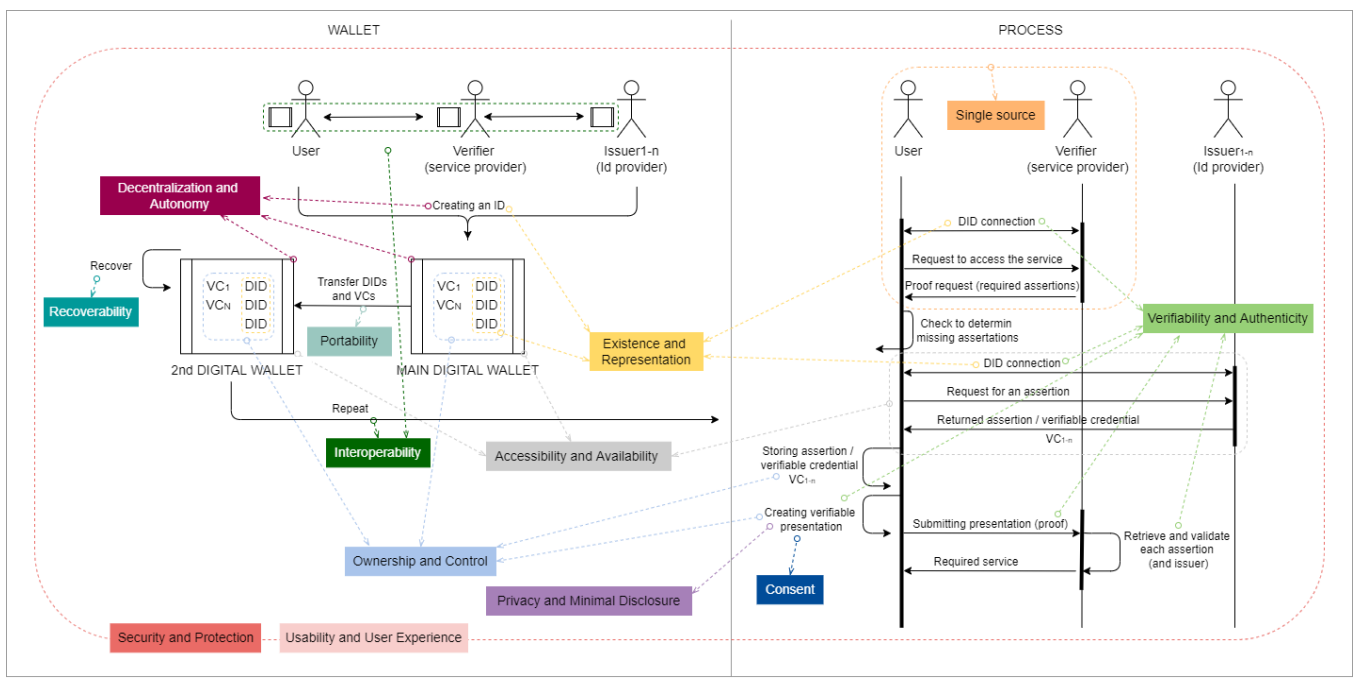

**FIGURE 3.** The general SSI process flow and connected properties.

### D. ROLES AND PROPERTIES

Each entity can act in various roles, e.g., identity holder, issuer, or verifier, depending on the situation. For example, (i) When an entity wants to access a particular service, he/she acts as an **identity holder** as he/she has to authenticate and provide credentials to prove that he/she meets the required conditions for service access. The same entity can also act as (ii) An **issuer** when he/she provides an opinion about a service or, (iii) A **verifier** when he/she requests a credit rating or proof of a service quality from another party, i.e., service provider. Thus, the sensitivity, importance, and context of each property vary according to the role an entity plays in a given situation. Based on the definitions and body of knowledge on this topic, we elaborate on and highlight the importance of properties towards the roles as follows (a summarized matrix is presented in Table 4). Each entity (Natural Person or Legal Entity) in the SSI ecosystem is an identity holder, while the roles of issuer and verifier usually belong to Legal Entities. Although a Natural Person can also act in the mentioned roles, as described above.

As a holder, the ability to create as many identities as required for digital interactions is crucial (**Existence and Representation**), since it allows an individual to interact and present him/herself in different contexts (e.g., professional and personal use cases). SSI allows him/her to create an identity, i.e., identifier and self-attested attributes, without the intervention of any intermediary. However, building a trusted identity still requires the involvement of issuers, who attest to certain identity claims in the form of VC. VCs, DIDs, and associated cryptographic keys are stored in the holder's digital wallet and are completely under his/her control, as he/she is responsible for provisioning their identity through his/her digital wallet (**Ownership and Control**). Moreover, from the holder's perspective, authentication and authorization without the direct involvement of third-party identity providers are essential, as it allows him/her to preserve their autonomy (**Decentralization and Autonomy**) in the digital environment. Additionally, being a single source of the truth regarding his/her identity (**Single Source**) is crucial for maintaining control and avoiding data transmission without his/her knowledge or consent (**Consent**), while he/she can preserve their privacy by exposing only the minimum data set required for successful completion of a particular interaction (**Privacy and Minimal Disclosure**).

From a security perspective, it is imperative that he/she trusts the technology and involved entities, and has an overall sense of protection when transmitting and storing identity data (**Security and Protection**). It is crucial that he/she feels safe and trusts that data breaches, abuses, errors and other security threats are well taken care of.

Since the identity holder is responsible for his/her digital identity (generation, storage, management, acquisition and distribution of identity data), it is paramount that (i) His/her wallet not only supports the mentioned operations, but is easy to use and offers a good user experience (**Usability and User Experience**), (ii) He/she has continuous access to the identity data in the wallet, and can obtain identity attributes instantly from issuers (**Accessibility and Availability**), (iii) He/she can restore his/her identity in case of problems (**Recoverability**), (iv) He/she can transfer identity data from one device or wallet to another (**Portability**), and (v) He/she has an overview of all the established connections, obtained credentials, shared data and all interactions (**Transparency**). It is also significant that the holder's identity is available as long as he/she requires it (**Persistence**), so he/she can use it in interactions with various entities and systems, not only within a particular domain (**Interoperability**). The cost of participating in the SSI ecosystem also affects the identity holders' decision to engage and use the technology greatly (**Cost**), whereby, from his/her point of view, it includes mainly digital wallet and transaction costs, as well as learning and required effort. On the other hand, the properties (i) **Decentralization**, (ii) **Standard**, and (iii) **Compatibility with legacy systems** are not that evident to the identity holders, while (iv) **Verifiability and Authenticity** are noteworthy from the context of ability to prove his/her identity reliably, but not from the verification context, which is in the domain of the verifiers.

From the perspective of issuers and verifiers, the properties related to authentication, authorization, issuance of verifiable credentials (issuer), provision of services (verifier), and ensuring compliance with the GDPR are paramount. To ensure the latter, entities in both roles must obtain consent from identity holders for activities such as data collection, processing, storage, etc. (**Consent**). Minimal data disclosure is also essential from the perspective of GDPR compliance, as it forces both parties to obtain only the minimum amount of data relevant for a single transaction. Thus, issuers only store and process data necessary for issuing credentials, while verifiers only acquire and process data required for provisioning a particular service (**Privacy and Minimal Disclosure**). Similar to the role of the identity holder, trust in the security of the technology is also critical for the issuer and verifier. Authentication is a critical step for both roles, to prevent misuse and minimize security threats. Issuers must authenticate identity holders before attesting to identity attributes and issuing verifiable credentials, and verifiers must authenticate and authorize identity holders to grant them access to the service. Verifiers must be able to determine (i) The issuer of the credential, (ii) That the credential has not been tampered with, and (iii) That it has not expired nor been revoked (**Security and Protection**, **Verifiability and Authenticity**). **Usability and User Experience**, as well as **Cost**, are influential for all roles. It is crucial that, in addition to the above, wallets support the entire process of (i) Issuing a verifiable credential (e.g., managing VC schemes, issuing, and offering VCs, etc.), and (ii) Verification of presentation (e.g., scheme compliance check, issuer, and credential validity check, etc.). Similar to the identity holder, verifiers and issuers are also affected by costs. Additional implementation or integration costs must be considered in the case of Legal Entities who need to integrate SSI into existing systems. On the other hand, (i) **Existence and Representation**, (ii) **Decentralization and Autonomy**, (iii) **Ownership and Control**, (iv) **Single Source**, (v) **Accessibility and Availability**, (vi) **Recoverability**, (vii) **Persistence**, (viii) **Portability**, (ix) **Interoperability**, are paramount properties when the entity acts as the identity holder, while they are not that significant when the entity acts as the issuer or verifier.

### E. BLOCKCHAIN AND PROPERTIES
As already mentioned, the SSI paradigm emerged with the rise of blockchain technology, as some blockchain properties can facilitate the exchange of data and build trust between entities without the need for intermediaries [3],

TABLE 4. The importance of properties for different roles.

| Role | Properties |
|---|---|
| Holder | Existence and Representation, Decentralization and Autonomy, Ownership and Control, Single Source, Accessibility and Availability, Recoverability, Persistence, Portability, Interoperability |
| Holder Issuer Verifier | Privacy and Minimal Disclosure, Consent, Security and Protection, Usability and User Experience, Transparency, Cost |
| Issuer Verifier | Verifiability and Authenticity, Compatibility with legacy systems |

[26]. Blockchain essentially provides a decentralized, immutable, append-only, trustless registry hosted on an open distributed network and presents a good foundation for SSI implementation [19] due to its decentralized nature (**Decentralization**) and default properties, such as (i) decentralized consensus protocol, (ii) immutability and irreversibility, (iii) data persistence and provenance, (iv) distributed data control, (v) accountability and transparency [5], [6].

It can serve as the Decentralized Public-Key Infrastructure (DPKI) [27], i.e., Identifier Registry, and acts as a replacement for the centralized registration authority (CA) in traditional Identity Management (IdM) systems, where the link is maintained, i.e., mapping between an identifier and an authentication method (public key) [6]. This, in the majority of SSI cases, is facilitated through the anchoring of DID documents on a publicly available blockchain ledger. The public permissionless blockchain allows entities to create and register as many identifiers as required (**Existence and Representation, Persistence**) without the intervention of a third party. However, the same applies to private or public permissioned blockchains, although, these require the participation of an administrative third party to enable writing operations. Identifiers and corresponding cryptographic keys are fully controlled and managed directly by the identity holders (**Control**), while identifier ownership can be proven with the corresponding private keys stored in entities' wallets (**Verifiability and Authenticity**), and verified with the help of the corresponding public keys accessible on the ledgers. In addition, blockchain can store the cryptographic proofs (hash) of all the associated credentials [6]. This allows entities to verify the integrity of the received credential by comparing it to the stored hash on the blockchain [5] (**Verifiability and Authenticity**). Because of its timestamping property, it can serve as a mechanism that can provide evidence regarding interaction between entities (e.g., proof of request/issuance/acceptance/revocation of VC, proof of request/sharing/acceptance/withdrawal of VP, etc.). Claim and signature modifications can be prevented by hashing evidence and storing it on the ledger (**Protection, Verifiability and Authenticity**) [6], [26]. Even though DIDs and DID documents of entities are being stored on the ledgers, the common rule is that these do not hold any personal or privacy concerning information about the user (**Privacy and Minimal Disclosure**).

Although blockchain technology encompasses some of the desired properties of SSI, and although it has initiated and accelerated the development of decentralized identity, it is not always necessary for implementation, as (i) Distributed Ledger Technology (DLT) or other forms of decentralized systems can be used as DPKIs instead of blockchains, and (ii) There are also DID methods [28], such as DID:peer and DID:key, which allow establishing DID connections and exchanging data completely outside decentralized registers [29]. Furthermore, the storage of DID documents does not happen automatically when a DID is generated (i.e., it also depends on the DID method) and, as such, is controllable (**Consent**) by the user.

## IV. EXPERT VALIDATION
### A. METHODOLOGY

The set of eighteen properties and their classification obtained and presented in the previous Section was used in the questionnaire, which was the main research method used in this study.

A two-part questionnaire about Self-Sovereign Identity was conducted from 1st April to 21st May 2021 among experts in the field of Decentralized and Self-Sovereign Identity Management. The experts were chosen carefully through dedicated projects, groups, organizations, and/or forums, dealing with the topic of Decentralized and/or Self-Sovereign Identities. Its aim was to gain a broader insight into the perception of the SSI concept and its properties. Moreover, it was used to validate the final set of properties and the categorization by experts.

**Objectives:** The goal of the survey was to investigate if the identified properties, their naming, and definition, as well as proposed classification, were consistent with the opinion of the respondents. Moreover, the goal was to determine the perceived level of importance of the identified properties, determine the most and least important properties, and provide their classification according to perceived relevance and scope. Furthermore, while obtaining experts' opinions, the goal was to identify additional concerns, inconsistencies, misunderstandings, and properties that might have been overlooked.

**Questionnaire structure:** The questionnaire consisted of two parts. The first part was dedicated to obtaining demographic data, in order to provide an appropriate profile of respondents, while the second part dealt with SSI properties and their classification. For each property, a definition (Table 1) was given at the beginning to provide the context and proper understanding. It was followed by three questions related to the level of importance and classification. With the first one, we were trying to measure the perceived level of importance of each property with a Likert scale consisting of the following Likert items (i) Not impor-tant (Irrelevant), (ii) Slightly important (Unnecessary), (iii) Moderately important (Useful), (iv) Important (Desirable), (v) Very important (Mandatory). The second question was addressing classification, as we were trying to determine if our classification (Table 3) was appropriate or not. Thus, the respondents had the opportunity to agree with the proposed classification or select another category (Privacy, Security, Usability and UX, Adoption and Sustainability, Controllability, Flexibility, Other). Moreover, the respondents were able to choose more of the categories listed or propose a new one. The last, open-ended question allowed the respondents to express their opinions, concerns, ask questions, or leave comments regarding a change in categorization, naming, or definition suggestion. After the first set of questions related to a specific property, a general question followed in which respondents had to choose the top five properties that they find the most important for the field of SSI.

**Participants and procedures:** The questionnaire was designed using an online surveying tool (1ka.si) and was available between 1st April and 21st May 2021. Beforehand we had identified organizations and expert working groups that focus on the development and standardization of the fundamental elements for establishing the SSI ecosystem, as well as projects that use SSI technology stacks. Thus, the experts we contacted and sent the invitation to fill out the questionnaire were identified and selected carefully from these organizations, groups, and projects, namely, the Decentralized Identity Foundation (DIF), the W3C Verifiable Credentials Working Group, the W3C Decentralized Identifier Working Group, European Blockchain Service Infrastructure (EBSI) projects, European Self-Sovereign Identity Framework Lab (eSSIF-Lab), some H2020 and similar cross-border projects like DE4A, mGov4EU and Kraken, some research laboratories and standardization groups on the topic of Blockchain technology, etc. Therefore, the questionnaire was sent to experts in the field of Decentralized and Self-Sovereign Identity Management via email, and was also posted on online platforms in relevant groups to increase its reach. Moreover, the appropriate profile of respondents was guaranteed by employing an extra set of questions regarding demographic data addressing respondents' work experience, job position, the field of work, experience in the field of IdM and SSI, and place of residence. That approach ensured that they had at least some experience in the field of Decentralized and Self-Sovereign Identity Management.

Forty-four respondents participated in the survey. Among them, 12 (27.3%) dropped out before answering questions regarding SSI and were excluded from the analysis. The survey was answered partially by 5 (11.4%) and fully by 27 (61.4%), 32 (72.7%) respondents in total. Thus, the latter were taken into account accordingly.

The majority of the respondents had IT (11 respondents - 34.4%) and research (10 respondents - 31.3%) related positions, followed by business-related (5 respondents - 15.6%) ones. More than half of the respondents (18 respondents, 56.3%) work in the IT field while the remainder work in the fields of Science, Healthcare, Education, Government and Public Service, Business, Sales, Management, Agriculture, and Retail.

Regarding the number of years of experience in the field of IdM, 6 (18.8%) respondents had less than a year of experience, while the majority (13 respondents - 40.6%) had from 1 to 5 years of experience. Four (12.5%) from 6 to 10 years, five (15.6%) from 11 to 20 years and four (12.5%) more than 20 years of experience in IdM. Among them, many respondents who are involved actively in the standardization of SSI were interested in the results of the study. Therefore, we agreed to inform them individually, as well as their offices generally, and thus establish a quality loop to encourage additional research and standardization on this topic.

More details about the respondents can be found in the Appendix, where we provide a visualization of data about the respondents' work experience, job positions, the field of work, experience in the field of IdM, and place of residence (Figure 7 - 11).

**Limitations:** The study was limited to 18 properties that were included in the questionnaire and presented in Section III-A. It was also limited to the classifications presented in Section III-B. However, an additional category, namely Flexibility, was offered, to verify whether Ferdous' proposed categorization [5] was more appropriate than ours. Thus, the respondents could choose between all the proposed categories, not just ours. Moreover, participants were able to suggest new categories, and/or choose multiple categories. The number of respondents that took part in the questionnaire (32) presents another limitation. Therefore, a larger number of respondents would mean a greater validity of the results and a greater possibility of generalization. Nevertheless, as the field of SSI is a young research field, we were striving to receive truly experts in it, thus not forcing the increase of respondents by broadening the experts' scope, although further discussion and research will be needed to be able to offer bulletproof/solid classification. Therefore, we defined the results towards the classification.

### B. RESULTS AND DISCUSSION
1) Perceived level of importance

Concerning the perceived level of importance (Table 5), the average values for most properties are above the value of 4.00 (varying between 4.00 and 4.86), which means that most consider these properties to be either important (desirable) or very important (mandatory). The exceptions are Cost (AVG = 3.96) and Compatibility with legacy systems (AVG = 3.67), with an average value of less than 4.00. Cost is perceived mostly as an important (desirable) or very important (mandatory) property, while Compatibility with legacy systems is perceived as either important (desirable) or moderately important (useful).

TABLE 5. Importance level of identified SSI properties versus the top five most important properties chosen.

| Property | Top 5 | 1 | 2 | 3 | 4 | 5 | AVG | SD | Rank* | Rank** |
|---|---|---|---|---|---|---|---|---|---|---|
| Security and Protection | 13 / 48.15% | 0 / 0.00% | 0 / 0.00% | 1 / 3.45% | 2 / 6.90% | 26 / 89.66% | 4.86 | 0.44 | $1^{st}$ | $5^{th}-6^{th}$ |
| Verifiability and Authenticity | 15 / 55.56% | 0 / 0.00% | 0 / 0.00% | 1 / 3.57% | 4 / 14.29% | 23 / 82.14% | 4.79 | 0.50 | $2^{nd}$ | $2^{nd}-3^{nd}$ |
| Privacy and Minimal Disclosure | 16 / 59.26% | 0 / 0.00% | 0 / 0.00% | 1 / 3.45% | 5 / 17.24% | 23 / 79.31% | 4.76 | 0.51 | $3^{rd}$ | $1^{st}$ |
| Standard | 8 / 29.63% | 0 / 0.00% | 0 / 0.00% | 2 / 7.41% | 7 / 25.9% | 18 / 66.67% | 4.59 | 0.64 | $4^{th}$ | $7^{th}$ |
| Consent | 7 / 25.93% | 0 / 0.00% | 1 / 3.45% | 2 / 6.90% | 6 / 20.69% | 20 / 68.97% | 4.55 | 0.78 | $5^{th}$ | $8^{th}-9^{th}$ |
| Recoverability | 4 / 14.81% | 0 / 0.00% | 0 / 0.00% | 2 / 7.41% | 9 / 33.33% | 16 / 59.26% | 4.52 | 0.64 | $6^{th}$ | $11^{th}-13^{th}$ |
| Ownership and Control | 15 / 55.56% | 2 / 6.90% | 0 / 0.00% | 1 / 3.45% | 5 / 17.24% | 21 / 72.41% | 4.48 | 1.09 | $7^{th}$ | $2^{nd}-3^{nd}$ |
| Portability | 4 / 14.81% | 0 / 0.00% | 0 / 0.00% | 4 / 14.81% | 10 / 37.04% | 13 / 48.15% | 4.33 | 0.73 | $8^{th}$ | $11^{th}-13^{th}$ |
| Accessibility and Availability | 3 / 11.11% | 0 / 0.00% | 1 / 3.70% | 4 / 14.81% | 7 / 25.93% | 15 / 55.56% | 4.33 | 0.88 | $9^{th}$ | $14^{th}-15^{th}$ |
| Persistence | 0 / 0.00% | 0 / 0.00% | 1 / 3.70% | 4 / 14.81% | 7 / 25.93% | 15 / 55.56% | 4.33 | 0.88 | $10^{th}$ | $17^{th}-18^{th}$ |
| Interoperability | 13 / 48.15% | 0 / 0.00% | 0 / 0.00% | 3 / 11.11% | 13 / 48.15% | 11 / 40.74% | 4.30 | 0.67 | $11^{th}$ | $5^{th}-6^{th}$ |
| Usability and User Experience | 7 / 25.93% | 1 / 3.70% | 0 / 0.00% | 0 / 0.00% | 15 / 55.56% | 11 / 40.74% | 4.30 | 0.82 | $12^{th}$ | $8^{th}-9^{th}$ |
| Transparency | 4 / 14.81% | 1 / 3.70% | 0 / 0.00% | 2 / 7.41% | 12 / 44.44% | 12 / 44.44% | 4.26 | 0.90 | $13^{th}$ | $11^{th}-13^{th}$ |
| Existence and Representation | 5 / 18.52% | 2 / 6.25% | 1 / 3.13% | 1 / 3.13% | 11 / 34.38% | 17 / 53.13% | 4.25 | 1.11 | $14^{th}$ | $10^{th}$ |
| Decentralization and Autonomy | 14 / 51.85% | 3 / 10.34% | 0 / 0.00% | 4 / 13.79% | 7 / 24.14% | 15 / 51.72% | 4.07 | 1.28 | $15^{th}$ | $4^{th}$ |
| Single source | 1 / 3.70% | 2 / 6.90% | 1 / 3.45% | 5 / 17.24% | 8 / 27.59% | 13 / 44.83% | 4.00 | 1.20 | $16^{th}$ | $16^{th}$ |
| Cost | 3 / 11.11% | 1 / 3.70% | 2 / 7.41% | 4 / 14.81% | 10 / 37.04% | 10 / 37.04% | 3.96 | 1.09 | $17^{th}$ | $14^{th}-15^{th}$ |
| Compatibility with legacy systems | 0 / 0.00% | 0 / 0.00% | 1 / 3.70% | 10 / 37.04% | 13 / 48.15% | 3 / 11.11% | 3.67 | 0.73 | $18^{th}$ | $17^{th}-18^{th}$ |

1 = Not important (Irrelevant); 2 = Slightly important (Unnecessary); 3 = Moderately important (Useful); 4 = Important (Desirable);
5 = Very important (Mandatory); AVG = Average; SD = Standard Deviation
** = Rank based on the average level of importance [1-5], taking into account Standard Deviation
*** = Rank based on the 5 most important properties
20% ≤ ■ < 50% ≤ ■ < 70% ≤ ■
■ Most important properties ■ Least important properties

The majority, precisely more than half of the respondents, consider the following properties to be very important (mandatory): Security and Protection (26 - 89.66%), Verifiability and Authenticity (23 - 82.14%), Privacy and Minimal Disclosure (23 - 79.31%), Ownership and Control (21 - 72.41%), Consent (20 - 68.97%), Standard (18 - 66.67%), Recoverability (16 - 59.26%), Persistence (15 - 55.56%), Accessibility and Availability (15 - 55.56%), Existence and Representation (17 - 53.13%) and Decentralization and Autonomy (15 - 51.72%), while, Portability, Interoperability, Transparency, Cost, Single source, Usability and User Experience are almost equally distributed between being important and very important.

In general, most respondents believe that the identified properties are moderately important (useful), important (desirable), or very important (mandatory). However, there were some negligible outliers. Thus, some variability may be observed. The largest deviations occurred in the perceived level of importance of the properties Decentralization and Autonomy (SD = 1.28), Single source (SD = 1.20), Existence and Representation (SD = 1.11), Cost (SD = 1.09), and Ownership and Control (SD = 1.09), while the smallest deviation was at Security and Protection (SD = 0.44), Verifiability and Authenticity (SD = 0.50), Privacy and Minimal Disclosure (SD = 0.51).

According to the average value (AVG), taking into ac-count the Standard Deviation (SD), the properties can be arranged in order according to the perceived level of importance, as shown in Table 5 (Rank*), despite the extremely small differences between the average values. Thus, the five most important properties are (i) Security and Protection (AVG = 4.86, SD = 0.44), (ii) Verifiability and Authenticity (AVG = 4.79, SD = 0.50), (iii) Privacy and Minimal Disclosure (AVG = 4.76, SD = 0.51), (iv) Standard (AVG = 4.59, SD = 0.64), and (v) Consent (AVG = 4.55, SD = 0.78).

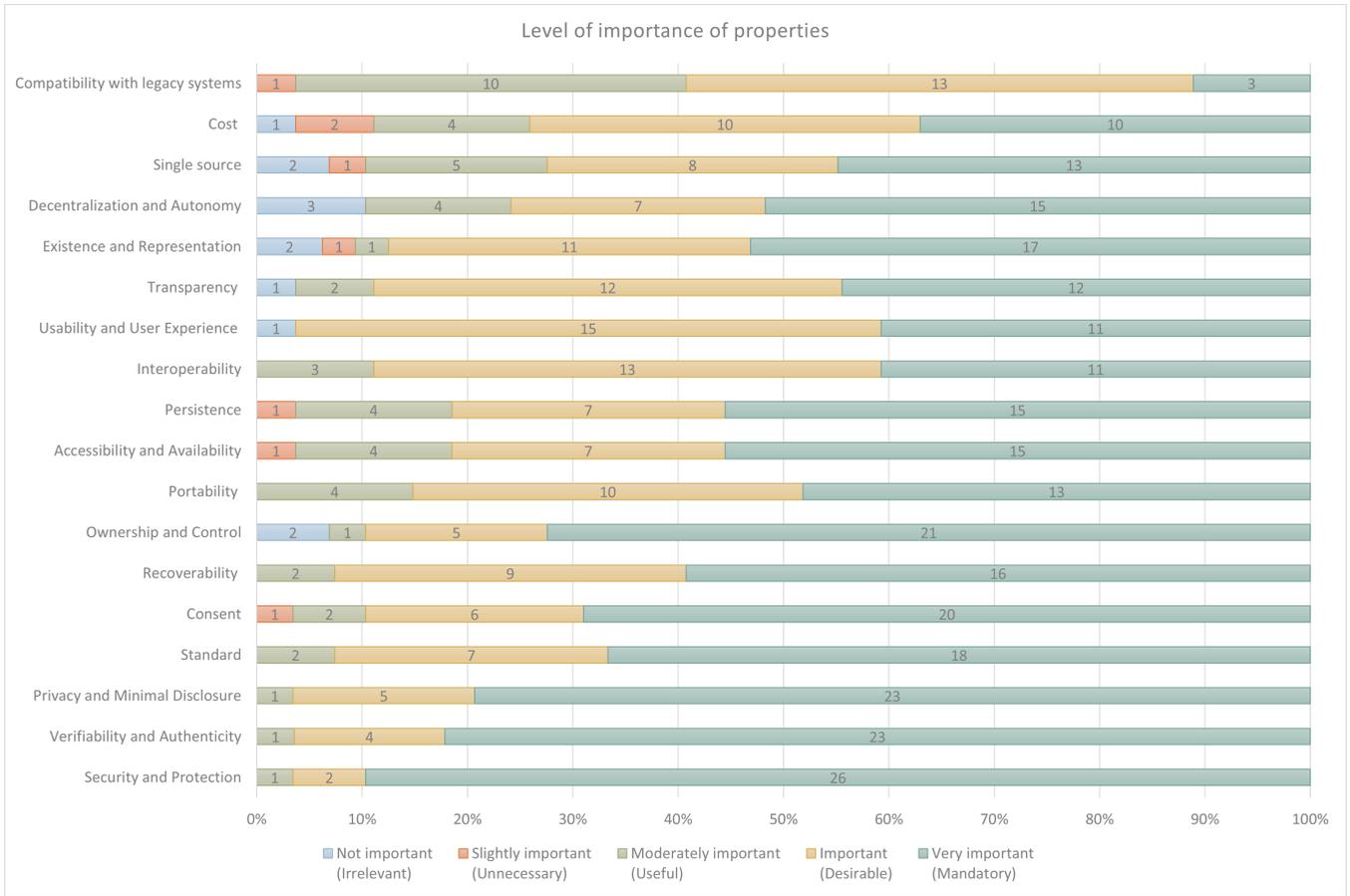

FIGURE 4. Perceived level of importance of properties.

Properties can be also ranked according to a selection of five properties that the respondents find most important (i.e. the Top 5) for SSI systems. The sequence is displayed in Table 5 (Rank**). Therefore, the most important properties according to this metric are (i) Privacy and Minimal Disclosure (16 - 59.26% respondents), (ii) Verifiability and Authenticity (15 - 55.56% respondents), (iii) Ownership and Control (15 - 55.56% respondents), (iv) Decentralization and Autonomy (14 - 51.85% respondents), (v) Security and Protection (13 - 48.15% respondents) and (vi) Interoperability (13 - 48.15% respondents), followed by Standard and Consent, that are among the top 5 properties of Rank*. The most and least important properties, as well as their intersection, are presented in Figure 5.

A high correlation between both rankings can be observed ($\rho = 0.6$). While there is no deviation between the ranking of four properties (Verifiability and Authenticity, Transparency, Single source, Compatibility with legacy systems), minor differences between both ranks of other properties exist, and can be observed in Table 5. The inconsistency in the ranking is extremely prominent for the property Decentralization and Autonomy. The property ranks 15th in terms of the first ranking (Rank*) and fourth in the second ranking (Rank**), with an astonishing 11 place difference. This could be related to the already mentioned variability (SD = 1.28). The property was ranked among the five most important by 51.85% of respondents. On the other hand, 10.34% of respondents believe that it is not important (irrelevant), and 13.79% that it is only moderately important (useful). Hence, according to the first ranking based on the perceived level of importance, the property Decentralization and Autonomy falls into the five least important properties. On the other hand, according to the second ranking, based on the selection of the most important properties, it is considered as being one of the five important properties of SSI. Regardless, minor differences between the average values (SD = 0.31) suggest that, in general, all the properties are important to some extent, and must be considered before SSI system implementation.

2) Classification

Regarding classification, the majority of the respondents agreed with the proposed categorization of individual properties presented in Table 3. All agreed with the categorization of Ownership and Control, Privacy and Minimal Disclosure, Security and Protection, Verifiability and Authenticity, Usability and User Experience, Standard, Portability, Interoperability, Compatibility with legacy systems, and Cost, while 96.30% of respondents concurred with the proposed categorization of the rest of the properties (Table 6), meaning that one individual did not agree with properties' placements. Overall, these outliers represent 4 individuals (out of 32 respondents) who chose a combination of the remaining categories instead of the proposed category.

This indicates confirmation or agreement with our categorization. However, the respondents chose several categories on average (AVG = 1.84, SD = 1.14) instead of one, which is not surprising, since respondents had the option to change the proposed category by choosing one or several categories or propose a new one.

The greatest variability in the chosen categorization was observed in Existence and Representation (AVG = 2.25, SD = 1.48), Decentralization and Autonomy (AVG = 2.52, SD = 1.33), Verifiability and Authenticity (AVG = 1.75, SD = 1.32), Transparency (AVG = 1.85, SD = 1.32), and Ownership and Control (AVG = 2.28, SD = 1.31), where the biggest discrepancies occurred. On the other hand, less variability and, thus, greater agreement, was detected in the categorization of the properties Accessibility and Availability (AVG = 1.59, SD = 0.80), Standard (AVG = 1.59, SD = 0.89), Compatibility with legacy systems (AVG = 1.52, SD = 0.94), Cost (AVG = 1.33, SD = 0.96), Interoperability (AVG = 1.67, SD = 0.96) and Persistence (AVG = 1.52, SD = 0.98), where more than 55.50% of respondents chose only one category.

The results reflect the nature of the properties that intertwine and complement each other, making their clear demarcation and categorization difficult. For some properties, the complementarity is particularly prominent, reflecting in several chosen classification categories. This is especially noticeable with the following properties, Recoverability (70.37%), Decentralization and Autonomy (68.97%), Consent (65.52%), Ownership and Control (62.07%), Existence and Representation (59.38%), and Privacy and Minimal Disclosure (51.72%), where more than half of the respondents chose several categories.

For the following properties: Security and Protection, Verifiability and Authenticity, Accessibility and Availability, Transparency, Standard, Persistence, Portability, Interoperability, Compatibility with legacy systems, and Cost, the choice of one category prevailed. More than 90% of respondents agreed with our categorization, while the choice of the remaining categories was less than 30%. For other properties, in addition to the proposed category, 30% or more respondents chose one or two additional categories, while the choice of the remaining categories was less than 30%. The distribution between chosen categories is as follows: Existence and Representation (Controllability - 96.88%, Privacy - 53.13%), Decentralization and Autonomy (Controllability - 96.55%, Privacy - 55.17%, Security - 34.48%), Ownership and Control (Controllability - 100.00%, Privacy - 48.28%, Security - 41.38%), Privacy and Minimal Disclosure (Controllability - 34.48%, Privacy - 100.00%), Single source (Controllability - 31.03%, Privacy - 96.55%), Consent (Controllability - 51.72%, Privacy - 96.55%), Recoverability (Security - 40.74%, Usability and User Experience - 96.30%, Adoption and Sustainability - 40.74%), Usability and User Experience (Usability and User Experience - 100.00%, Adoption and Sustainability - 37.04%).

The precise distribution of categories to which respondents believed each property belongs can be observed in Table 6 and is presented visually in Figure 6.

## V. DISCUSSING PROPERTIES

The definitions presented in Table 1 were abbreviated as much as possible intentionally, due to their use in the questionnaire. In this Section, we want to enhance them in accordance with the results and comments of the respondents, as we obtained valuable insights from experts in the fields of IdM and SSI, expressing their concerns and possible misunderstandings of an individual property.

Below you can find properties that have either (i) Required an extension or correction of the definition, or (ii) Have required additional discussion reflecting the obtained results, while other properties are omitted intentionally.

Definitions are italicized in quotation marks. Changes are bold, while the parts that need to be removed are crossed out. Afterwards, a discussion was added in normal fonts.

**Existence and Representation:** "Entities must have an independent existence. They should be able to create as many identities as required without the intervention of a third party."

Entities should be allowed to generate/create identifiers (DIDs) for each interaction separately. This increases controllability, flexibility, and privacy, as multiple identifiers reduce linkability while enabling entities to present themselves differently in different contexts. Security can also be increased, since dependency on trusted third parties is reduced. However, this depends on the involved entities. Moreover, entities can self-assert an unlimited number of identities (verifiable presentations) without any third party involvement. However, trusted third parties are needed for issuing verifiable credentials and validation of "true" identities, reducing fraud and impersonation. Note that identifier must not be confused and equated with identity.

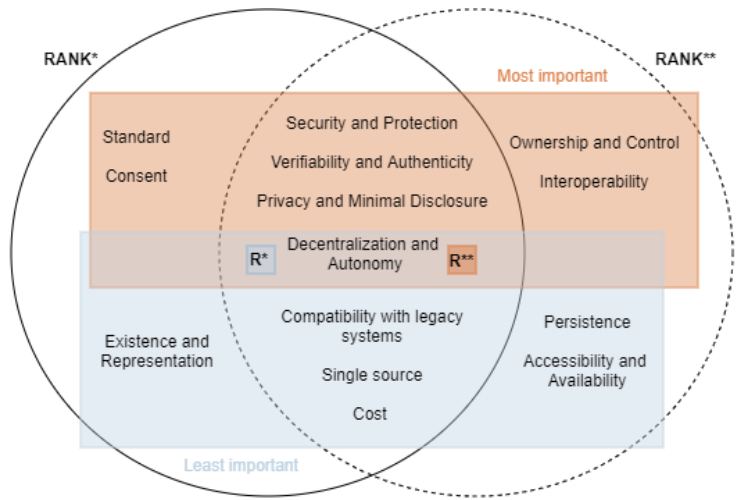

**FIGURE 5.** The most and least important properties of both rankings.

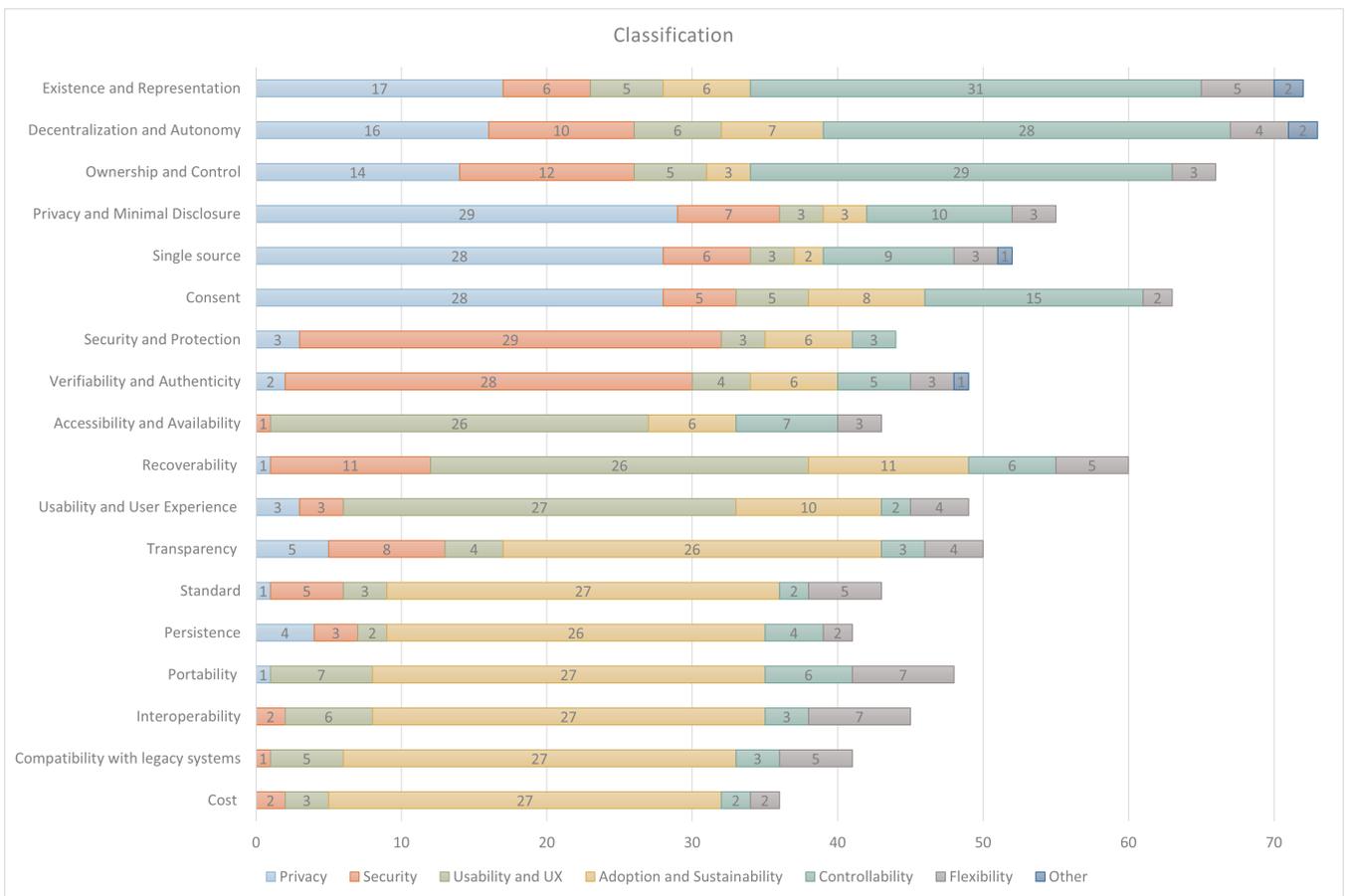

**FIGURE 6.** Classification of properties based on respondents' opinions.

TABLE 6. Categorization of SSI properties according to the proposed categories.

| Property | C1 | C2 | C3 | C4 | C5 | C6 | C7 | AVG | SD | 1≤ |
|---|---|---|---|---|---|---|---|---|---|---|
| Existence and Representation | 31 96.88% | 17 53.13% | 6 18.75% | 5 15.63% | 6 18.75% | 5 15.63% | 2 6.25% | 2.25 | 1.48 | 19 59.38% |
| Decentralization and Autonomy | 28 96.55% | 16 55.17% | 10 34.48% | 6 20.69% | 7 24.14% | 4 13.79% | 2 6.90% | 2.52 | 1.33 | 20 68.97% |
| Ownership and Control | 29 100.00% | 14 48.28% | 12 41.38% | 5 17.24% | 3 10.34% | 3 10.34% | 0 0.00% | 2.28 | 1.31 | 18 62.07% |
| Privacy and Minimal Disclosure | 10 34.48% | 29 100.00% | 7 24.14% | 3 10.34% | 3 10.34% | 3 10.34% | 0 0.00% | 1.90 | 1.21 | 15 51.72% |
| Single source | 9 31.03% | 28 96.55% | 6 20.69% | 3 10.34% | 2 6.90% | 3 10.34% | 1 3.45% | 1.79 | 1.01 | 14 48.28% |
| Consent | 15 51.72% | 28 96.55% | 5 17.24% | 5 17.24% | 8 27.59% | 2 6.90% | 0 0.00% | 2.17 | 1.23 | 19 65.52% |
| Security and Protection | 3 10.34% | 3 10.34% | 29 100.00% | 3 10.34% | 6 20.69% | 0 0.00% | 0 0.00% | 1.52 | 1.18 | 6 20.69% |
| Verifiability and Authenticity | 5 17.86% | 2 7.14% | 28 100.00% | 4 14.29% | 6 21.43% | 3 10.71% | 1 3.57% | 1.75 | 1.32 | 12 42.86% |
| Accessibility and Availability | 7 25.93% | 0 0.00% | 1 3.70% | 26 96.30% | 6 22.22% | 3 11.11% | 0 0.00% | 1.59 | 0.80 | 12 44.44% |
| Recoverability | 6 22.22% | 1 3.70% | 11 40.74% | 26 96.30% | 11 40.74% | 5 18.52% | 0 0.00% | 2.22 | 1.22 | 19 70.37% |
| Usability and User Experience | 2 7.41% | 3 11.11% | 3 11.11% | 27 100.00% | 10 37.04% | 4 14.81% | 0 0.00% | 1.81 | 1.24 | 13 48.15% |
| Transparency | 3 11.11% | 5 18.52% | 8 29.63% | 4 14.81% | 26 96.30% | 4 14.81% | 0 0.00% | 1.85 | 1.32 | 11 40.74% |
| Standard | 2 7.41% | 1 3.70% | 5 18.52% | 3 11.11% | 27 100.00% | 5 18.52% | 0 0.00% | 1.59 | 0.89 | 10 37.04% |
| Persistence | 4 14.81% | 4 14.81% | 3 11.11% | 2 7.41% | 26 96.30% | 2 7.41% | 0 0.00% | 1.52 | 0.98 | 7 25.93% |
| Portability | 6 22.22% | 1 3.70% | 0 0.00% | 7 25.93% | 27 100.00% | 7 25.93% | 0 0.00% | 1.78 | 1.09 | 11 40.74% |
| Interoperability | 3 11.11% | 0 0.00% | 2 7.41% | 6 22.22% | 27 100.00% | 7 25.93% | 0 0.00% | 1.67 | 0.96 | 10 37.04% |
| Compatibility with legacy systems | 3 11.11% | 0 0.00% | 1 3.70% | 5 18.52% | 27 100.00% | 5 18.52% | 0 0.00% | 1.52 | 0.94 | 8 29.63% |
| Cost | 2 7.41% | 0 0.00% | 2 7.41% | 3 11.11% | 27 100.00% | 2 7.41% | 0 0.00% | 1.33 | 0.96 | 4 14.81% |

C1 = Controllability; C2 = Privacy; C3 = Security; C4 = Usability and UX; C5 = Adoption and Sustainability; C6 = Flexibility; C7 = Other; AVG = The average number of chosen categories; SD = Standard Deviation of the chosen categories; 1≤ = More than one category chosen
30% ≤ ■ < 50% ≤ ■ < 90% ≤ ■
■ Agreement with our categorization

**Decentralization and Autonomy:** "Entities must have full autonomy over their identity data without relying on any third party (centralized system). They should be *capable of being* responsible for managing all operations related to their identity and data (creating, storing, updating, sharing, removing)."

While decentralization refers to the absence of central systems, autonomy refers to the management of identities (e.g., control over distribution/data disclosure/number of identities an entity possesses). Respondents noted a common misunderstanding of this principle. Thus, further discussion is needed at this point. In this definition, creation refers to identifiers and self-asserted credentials. As mentioned earlier, in order to attain a trusted identity, verifiable credentials still have to be obtained from third-party issuers. To prevent misunderstandings, we emphasize that an entity cannot issue itself, e.g., passports or other documents issued by government bodies or other institutions. It should also be emphasized that autonomy is not about individuals being completely independent of external third parties, but rather about being autonomous in creating identifiers and self-asserted credentials, and being autonomous in interactions with other parties. Therefore, after obtaining verifiable credentials from trusted issuers, entities can store them autonomously, generate verifiable presentations, and present them to third parties without issuers being aware of their usage, similar to the physical world.

Due to the above, it would make sense to separate autonomy and decentralization into two separate properties.

Another thing that should be pointed out is whether we should talk about full decentralization at all, as every credential issuer uses and relies on centralized systems and cannot operate without them. Therefore, decentralization should be about minimizing and removing the strict dependence on a third party, not eliminating it entirely.

**~~Ownership~~ and Control:** "Entities must ~~own and~~ control their digital identities and the involved data fully (e.g., self-asserted claims or claims provided by third parties, identifiers, encryption keys). They should be able to control the usage/sharing of their identity data and delegating control to autonomous agents and/or guardians of their choice."
We agree with the respondents that control and ownership should not be associated, since data are controlled by individuals, but owned by organizations (except self-asserted credentials). For instance, an education credential, driving license, passport, credit card, etc., are issued and owned by third-party institutions (issuers). However, the subject of the credentials should be able to control how those credentials are stored and shared, which increases privacy and security. SSI enhances control, not ownership, therefore, the property should be corrected accordingly.

**Privacy and Minimal Disclosure:** "Entities should be able to protect their privacy by utilizing selective disclosure and data minimization. They should be able to disclose the minimum amount of identity data required for any particular interaction."
However, respondents pointed out the subjectivity of this property, since the minimally required data set for a particular interaction depends on the perception of the participating parties and the risk of a particular transaction. Thus, some formalization/rules must be enforced. Minimal Disclosure is also connected with Security and Controllability, since entities are in control over which data they are willing to share with third parties.

**Single source:** "Entities should be the single source of truth regarding their identities. They should be able to create self-asserted claims, accumulate claims from third parties, *and control* and distribute them when required. Third parties should not be able to exchange entities' data without their knowledge and consent."
However, it should be noted that some exceptions and concerns exist. First of all, the statement is predominantly true for individuals and businesses, not for things. Secondly, in some cases, it is either preferred or required to involve a trusted third party for (help with) managing identity data (e.g., elderly, non-tech savvy people, minors, disabled people, pets, things, etc.) on behalf of the subject. Therefore, the subject of the credential is not always the controller. Thirdly, it is hard to enforce full copy protections and control the usage of data once a third party gathers it. Lastly, it is challenging to prevent governments and law enforcement from exchanging personally identifiable information (PII) about citizens and criminals. On the other hand, commercial third parties should abide by the stated objective, which should be reinforced by legalization. Otherwise, exploitation of PII is inevitable.

**Consent:** "Entities should be able to give deliberate and well-understood consent for the usage/sharing of their identity data (e.g., consenting to accept data related to their identity). *In addition, they should be able to withdraw/revoke that consent at a later date.* "
However, the statement is predominantly true for individuals and businesses, not for things. It affects privacy, allowing flexibility, and can influence the adoption by users positively.

**Security and Protection:** "Digital identity should be secure and well protected with reliable cryptographic mechanisms. Entities must be authenticated and authorized properly prior, to be able to use their digital identity. Any identity information must be transmitted/transferred via a secure channel to prevent cyber attacks."
Moreover, the rights of entities should be protected. Therefore, mechanisms that provide hard evidence regarding interaction should be available, with sufficient assurance about the identity of both parties. Thus, the rights of entities can be protected by employing the appropriate precautions: Protecting entities from being sued or from facing claims regarding interactions they had not taken part in or were forced/coerced to take part in. Thus, we believe Security and Protection should be separated into two properties, as Security is focusing on technology, while Protection is addressing the protection of entities and their rights.

**Recoverability:** "The identity must be robust enough to be recoverable."
However, concerns were expressed regarding potential security vulnerability. They are addressing the possibility of attackers obtaining secrets required for recovering identity information. Concerns are also related to the portability of identity information from one device to another. Beforehand risk assessment is required, as recoverability often introduces a security risk. On the other hand, this property can be omitted totally if obtaining verifiable credentials is so simplified that it does not impose an excessive burden on entities, thus recovering identity could be done by starting over.

**Usability and User Experience:** "The usability of agents and other identity system components should be maximized. User interfaces should allow entities to control, manage, and use their identities intuitively, reliably, and effectively. It should offer a consistent user experience, hide underlying complexity, and should be easy to use."
A system that is not easy to use and is not useful will not be used or adopted widely, while a poor user interface/user experience can also contribute to human errors affecting security.

**Persistence:** "Identities must be persistent, and should exist for at least as long as it is required by their owner. Longevity and the dynamic nature require firm separation between ~~identity~~ *identifiers* and their claims that can be modified or removed as appropriate."
As mentioned above, an identifier should not be equated with identity, since identity consists of an identifier and connected identity data, requiring a change in definition.

**Portability:** "Identities must be portable. Entities should be able to move or transfer their identity data securely to agents or systems of their choice. Portability ensures entities' control over their data, and improves persistence over time."
However, concerns were expressed regarding the potential risk of fraud. The respondents believed that it should not be allowed to duplicate/port identity data to multiple platforms, as portability could potentially enable identity sharing that would allow multiple entities to use the same identity in various places at the same time.

**Interoperability:** "Identities must be as widely usable/available as possible and not limited to a specific domain. ~~Global identities~~ *Interoperability* might increase persistence and identity autonomy, *and can best be achieved with standardization.*"
In the above definition, with the term "global identity" we were trying to suggest that entities can create identities that can be used anywhere and are not limited to a specific domain. The term was replaced, to prevent the misunderstandings identified with the questionnaire.

**Compatibility with legacy systems:** "Identity should be backward compatible with legacy identity systems to ensure quicker acceptance."
However, we agree with the respondent about legacy systems becoming obsolete and being replaced by SSI systems in the future. Therefore, we believe this property is useful, but definitely not mandatory.

**Cost:** "The cost of identity creation, management, and adoption should be minimized. *The benefits of SSI must substantially outweigh the costs, otherwise, adoption might be hindered.*"
Therefore, addressing transactions fees should be considered when using public blockchains, like bitcoin. To minimize costs, the majority of data should be stored off-chain, while specialized blockchains like Hyperledger Indy can be used for storing encrypted proof.

**Final set of properties:** Based on the results obtained from the questionnaire and based on the above analysis, we have prepared a final list of properties, their definition, and classification. The properties defined in Subsection III-A and their classification, provided in Subsection III-B, are adjusted according to the findings of the validation phase and are presented in Table **7**.

## VI. CONCLUSION

With the growing interest in decentralized technologies in academia and industry, the number of proposed decentralized identity solutions is increasing rapidly. However, they do not always comply with the criteria that an SSI system should possess. Furthermore, inconsistencies exist about the notion and importance of various SSI properties. Thus, this study aims to clarify any misunderstandings, and distinguish between the essential properties of SSI and properties that can be neglected, according to the perception of experts in the fields of Identity Management and SSI. In addition, properties are classified into five categories, and a general SSI process is presented, highlighting the process steps in which individual properties are important. Moreover, the importance of individual properties for various roles is analyzed, and the properties that can be satisfied by utilizing Blockchain technology are highlighted.

The results of our research show that the majority of the identified SSI properties are perceived as being important (desirable) or very important (mandatory), with an average value above 4.00 (varying between 4.00 and 4.86) by the experts in the fields of IdM and SSI. Security and Protection (AVG = 4.86), Verifiability and Authenticity (AVG = 4.79), Privacy and Minimal Disclosure (AVG = 4.76), Ownership and Control (AVG = 4.48) are considered as being mandatory. The exceptions (with an average below 4.0) are Cost (AVG = 3.96) and Compatibility with legacy systems (AVG = 3.67), which linger between being useful and desirable.

Regardless, we believe that, in the implementation of an SSI solution, it is necessary to strive to meet as many SSI properties as possible, while finding a balance between the properties, requirements, and needs of each system individually. As mentioned in Section II, meeting all the SSI properties is a challenging task. Most of the existing research that has analyzed systems, defined as SSI, do not fulfill all the properties, indicating that meeting all the properties is not always possible. Therefore, it is imperative to determine the essential properties that must be fulfilled to be labeled self-sovereign. At this point, our ranking and definition of the most important and least important properties come into play. According to our research, the most important properties are Security and Protection, Verifiability and Authenticity, Privacy and Minimal Disclosure, Ownership and Control, Interoperability, Standard, and Consent. While the least important are Compatibility with legacy systems, Single source, Cost, Persistence, Accessibility and Availability, and Existence and Representation.

Regarding classification, the majority (>95%) of the respondents agreed with the proposed categorization (Controllability, Privacy, Security, Usability and User Experi-ence, Adoption and Sustainability). However, several categories per property were chosen on average, reflecting the nature of the properties that intertwine and complement each other, making their clear demarcation and categorization difficult. This is especially evident in the categories with the greatest variability.

In addition to large overlaps between SSI properties, there is still ambiguity regarding the understanding of the concept. Respondents' comments indicate inconsistencies in the interpretation of individual properties. This is especially noticeable with the properties Existence and Representation, Decentralization and Autonomy, Ownership and Control, and Single source, where the greatest variability regarding the perceived level of importance is observed, as well as variability in their classification.

It should also be noted that the presented properties address primarily situations dealing with individuals. Therefore, future research should rethink and adapt principles to meet the needs and requirements of things and organizations/businesses. Moreover, it would be useful to determine which properties are eligible for each entity type (e.g., things, organizations/businesses, etc.) while highlighting similarities and differences. In addition, our future research might include improving this study by conducting an improved survey questionnaire that would (i) Include a larger set of respondents, (ii) Involve our final, refined set of SSI properties, and would also (iii) Address blockchains' importance in SSI implementation and (iv) Differentiate the SSI roles and their sensitivity to identified properties by employing an additional set of questions which would enable us to validate the analysis presented in Subsections III-D, III-E.

Additionally, it would also be useful to (i) Extend the results of existing studies [5], [8], [17]–[19] that have analyzed prototypes according to properties fulfillment by adding additional properties' identified in this study, and (ii) Analyze recent SSI prototypes to reinforce theoretical knowledge with practical. Thus, one of our potential further research directions can be the analysis of SSI prototypes according to the final set of properties.

TABLE 7: SSI final list of properties, their definitions and classification.

| Property | Definition | Classification |
| --- | --- | --- |
| Existence and Representation | Entities must have an independent existence. They should be able to create as many identities as required without the intervention of a third party. | Controllability, Privacy* |
| Decentralization | SSI systems should not rely on any third-party centralized system. | Controllability, Privacy*, Security* |
| Autonomy | Entities must have full autonomy over their identity data without relying on any third party. They should be capable of being responsible for managing all operations related to their identity and data (creating, storing, updating, sharing, removing). | Controllability, Privacy*, Security* |
| Control | Entities must control their digital identities and the involved data fully (e.g., self-asserted claims or claims provided by third parties, identifiers, encryption keys). They should be able to control the usage/sharing of their identity data and delegating control to autonomous agents and/or guardians of their choice. | Controllability, Privacy*, Security* |
| Privacy and Minimal Disclosure | Entities should be able to protect their privacy by utilizing selective disclosure and data minimization. They should be able to disclose the minimum amount of identity data required for any particular interaction. | Controllability*, Privacy |
| Single source | Entities should be the single source of truth regarding their identities. They should be able to create self-asserted claims, accumulate claims from third parties, and control and distribute them when required. Third parties should not be able to exchange entities' data without their knowledge and consent. | Controllability*, Privacy |
| Consent | Entities should be able to give deliberate and well-understood consent for the usage/sharing of their identity data (e.g., consenting to accept data related to their identity). In addition, they should be able to withdraw/revoke that consent at a later date. | Controllability*, Privacy |
| Security | Digital identity should be secure and well protected with reliable cryptographic mechanisms. Entities must be authenticated and authorized properly prior, to be able to use their digital identity. Any identity information must be transmitted/transferred via a secure channel to prevent cyber attacks. | Security |
| Protection | The rights of entities should be protected by employing the appropriate precautions. Therefore, mechanisms that provide hard evidence regarding interaction should be available, with sufficient assurance about the identity of both parties. | Security |
| Verifiability and Authenticity | Entities must be able to prove their identity reliably. They must provide verifiable proof of authenticity of digital identity data. Relying parties should be able to verify that digital identities are controlled by their owners and haven't been tampered with. | Security |
| Accessibility and Availability | Entities must have unrestricted access to their identity information. They must be able to retrieve claims and assertions (self-asserted or provided by a third party) that constitute their identity, and must be accessible and available from different platforms when required. | Usability and UX |
| Recoverability | The identity must be robust enough to be recoverable. | Usability and UX, Security*, Adoption and Sustainability* |
| Usability and User Experience | The usability of agents and other identity system components should be maximized. User interfaces should allow entities to control, manage, and use their identities intuitively, reliably, and effectively. It should offer a consistent user experience, hide underlying complexity, and should be easy to use. | Usability and UX, Adoption and Sustainability* |
| Transparency | The identity system and algorithms must be transparent enough for every involved entity. They should be free, open-source, well-known, and independent of any particular architecture. Entities should be well aware of all their partial identities and their corresponding interactions. | Adoption and Sustainability |
| Standard | Identities must be based on open Standards to ensure maximal portability, interoperability, and persistence. Entities should be represented, exchanged, secured, protected, and verified using open, public, and royalty-free Standards. | Adoption and Sustainability |
| Persistence | Identities must be persistent, and should exist for at least as long as it is required by their owner. Longevity and the dynamic nature require firm separation between identifiers and their claims that can be modified or removed as appropriate. | Adoption and Sustainability |

| Portability | Identities must be portable. Entities should be able to move or transfer their identity data to agents or systems of their choice securely. Portability ensures entities' control over their data and improves persistence over time. | Adoption and Sustainability |
| --- | --- | --- |
| Interoperability | Identities must be as widely usable/available as possible, and not limited to a specific domain. Interoperability might increase persistence and identity autonomy, and can be best achieved with standardization. | Adoption and Sustainability |
| Compatibility with legacy systems | Identity should be backward compatible with legacy identity systems to ensure quicker acceptance. | Adoption and Sustainability |
| Cost | The cost of identity creation, management, and adoption should be minimized. The benefits of SSI must substantially outweigh the costs, otherwise, adoption might be hindered. | Adoption and Sustainability |

The classification categories were proposed originally in Section III-B. Other listed categories, marked with *, were chosen additionally by more than 30% of the experts.

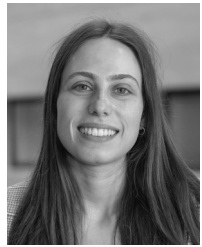

ŠPELA ČUČKO received a Master's degree in Informatics and Technology of Communication from the Faculty of Electrical Engineering and Computer Science at the University of Maribor, where she is currently pursuing a Ph.D. in Computer Science. She is currently a Young Researcher at the Institute of Informatics and a member of the Research and Development Group named Blockchain Lab:UM. Her current research interests include Digital, Decentralized, and Self-Sovereign Identities.

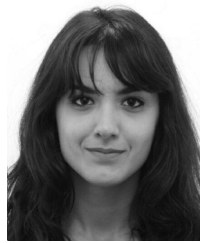

ŠEILA BEĆIROVIĆ received a B.Sc. degree in Computer Science and Informatics in 2017 and an M.Sc. degree in Computer Science and Informatics from the Faculty of Electrical Engineering at the University of Sarajevo, Bosnia and Herzegovina, in 2019. She currently works as a Teaching Assistant at the Faculty of Electrical Engineering at the University of Sarajevo. She is also a Full-Stack Software Engineer at InfoStudio d.o.o. Sarajevo. Her current research interests include Identity, Self-Sovereign Identity, Blockchain and its uses and applications.

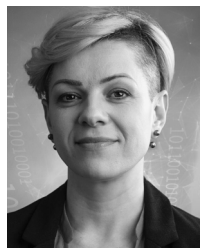

AIDA KAMIŠALIĆ is an Assistant Professor and researcher at the University of Maribor, Faculty of Electrical Engineering and Computer Science. She is a member of the Laboratory for Data Technologies and Blockchain Lab:UM team. She earned a PhD in Computer Science in 2014. Her research interests include Advanced Database Technologies, Blockchain Technology and Medical Informatics. She has co-authored several scientific papers published in renowned journals and a patent application in the field of Blockchain Technology. She has been involved in several applied and international projects. In January 2020 she was awarded the prestigious title of Female engineer of the year 2019 in Slovenia.

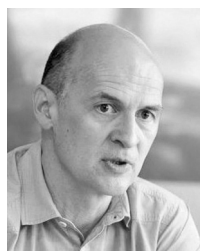

SAŠA MRDOVIĆ is a Professor at the Faculty of Electrical Engineering, University of Sarajevo. He teaches Computer Networks and Security courses. He defended his Ph.D. thesis on Intrusion Detection Systems at the Department for Computing and Informatics, in 2009. His main research interests include Digital Information Security, Digital forensics and IoT. Saša published three books on Networks and Security, and a number of papers in scientific journals and conference proceedings. He has been reviewing papers for various journals and conferences. Saša Mrdović also works on projects with industry and government in the area of Information Security. He holds a CISSP, security industry, Certificate.

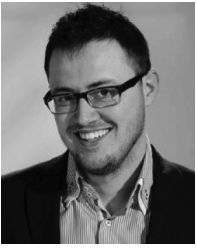

MUHAMED TURKANOVIĆ is currently an Associate Professor at the Institute of Informatics, University of Maribor (UM), Slovenia, Faculty of Electrical Engineering and Computer Science, where he is also the Head of Research and Development at the Blockchain Lab:UM, as well as the Head of the Slovene EDIH DIGI-SI, UM's coordinator of the H2020 project DE4A—Digital Europe for All etc. He was a Managing Director and a CTO of an IT Company, from 2013 to 2016. His current research interests include Advanced Database Technologies, Cryptography, Digital Identities and Blockchain.

## APPENDIX. RESPONDENT DEMOGRAPHICS

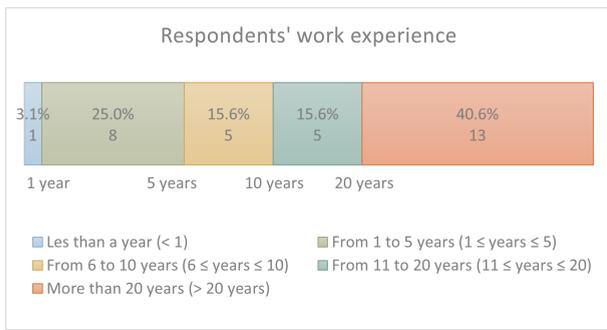

FIGURE 7. Respondents' work experience.

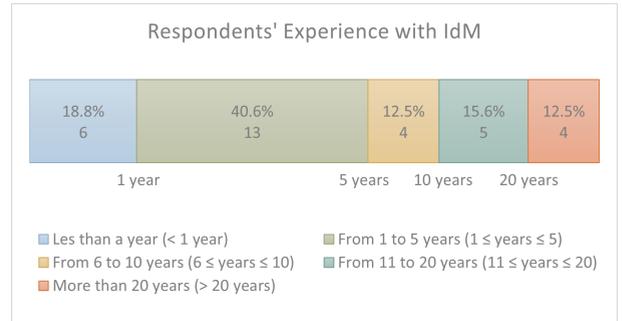

FIGURE 10. Respondents' experience in the field of Identity Management.

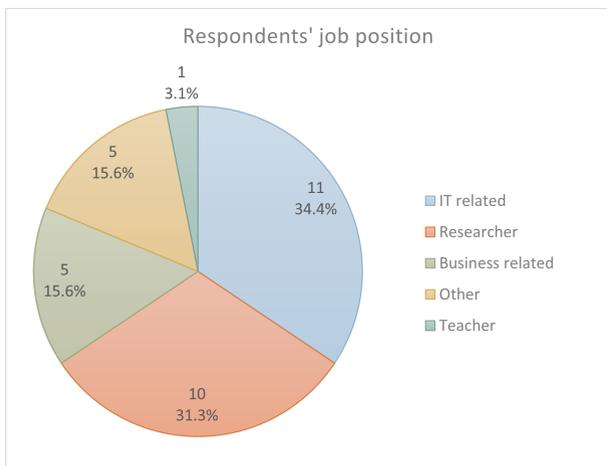

FIGURE 8. Respondents' job position.

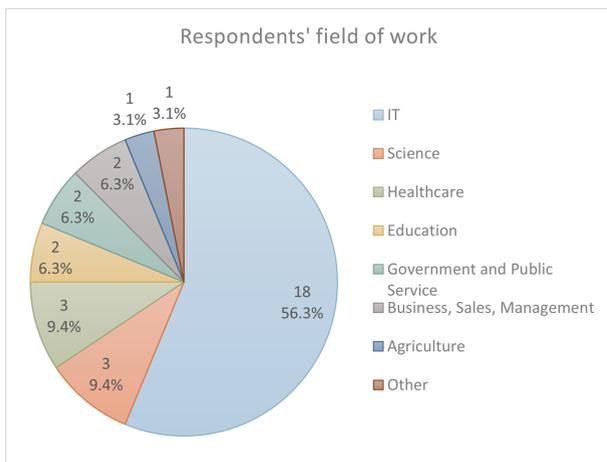

FIGURE 9. Respondents' field of work.

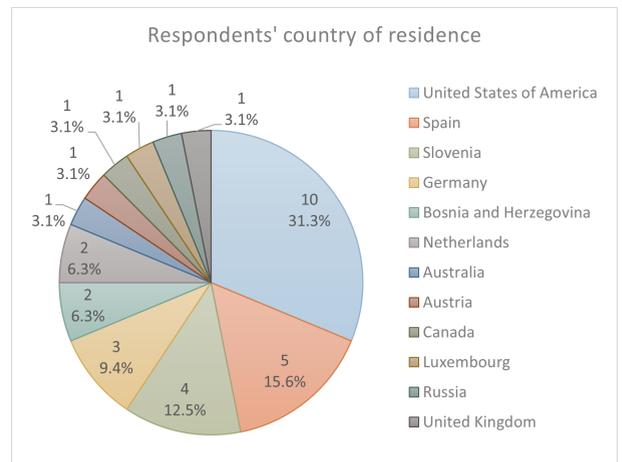

FIGURE 11. Respondents' country of origin.